# A Survey of Network-On-Chip Tools

Ahmed Ben Achballah
Dept. of Electrical Engineering
National Institute of Applied Sciences and Technology
Polytechnic School of Tunisia, LSA laboratory
Centre Urbain Nord, BP 676- 1080 Tunis Cedex, Tunisia.

Slim Ben Saoud
Dept. of Electrical Engineering
National Institute of Applied Sciences and Technology
Polytechnic School of Tunisia, LSA laboratory
Centre Urbain Nord, BP 676- 1080 Tunis Cedex, Tunisia.

*Abstract*—Nowadays System-On-Chips (SoCs) have evolved considerably in term of performances, reliability and integration capacity. The last advantage has induced the growth of the number of cores or Intellectual Properties (IPs) in a same chip. Unfortunately, this important number of IPs has caused a new issue which is the intra-communication between the elements of a same chip. To resolve this problem, a new paradigm has been introduced which is the Network-On-Chip (NoC). Since the introduction of the NoC paradigm in the last decade, new methodologies and approaches have been presented by research community and many of them have been adopted by industrials. The literature contains many relevant studies and surveys discussing NoC proposals and contributions. However, few of them have discussed or proposed a comparative study of NoC tools. The objective of this work is to establish a reliable survey about available design, simulation or implementation NoC tools. We collected an important amount of information and characteristics about NoC dedicated tools that we will present throughout this survey. This study is built around a respectable amount of references and we hope it will help scientists.

*Keywords—Embedded Systems; Network-On-Chip; CAD Tools; Performance Analysis; Verification and Measurement*

## I. INTRODUCTION

The insatiable market demands for more innovative technologies have induced a considerable evolution of the integration capacities in recent platforms. In fact, Semi-conductor industrials have offered, are offering and will continue to offer many powerful hardware chips. Gates scaling continue to fall down (40nm, 35nm and recently 28nm), also power consumption is decreasing and GHz working frequencies are increasing [1-2]. A chip with the last cited advantages will enlarge the intervention domain of engineers and many design issues can be solved because the computing power and the chip flexibility are enormous.

Like the hardware side of technology, the software side which is represented by the Computer Aided Design (CAD) tools was dramatically innovated. This includes modeling, simulation, synthesis and implementation tools. Also, new design flows have emerged like the CoDesign concept. Moreover many techniques were proposed by research community and some of them were adopted by industrials like High Level synthesis (HLS) or Model Based Design (MBD). All of these enhancements and novelist techniques share common purposes i) To increase the abstraction level of a desired design flow ii) To furnish preventive estimations for engineers at earlier stages of the design to ensure low cost fixes iii) To accelerate the design flow.

Traditionally, a SoC is composed by some processing elements (processors, dedicated Intellectual Properties (IPs), etc), few memory blocks and In/Out communication modules. Nowadays, the number of these On-Chip elements is extremely growing: This is a direct result of the innovations and advancements cited earlier [1, 3]. Recent platforms are often Multi-Processors SoC (MPSoC) with multiple functionalities and a lot of options. For example we can cite recent personal computers, video games, smart phones and tablets. However, the growth of the On-Chip elements has provoked new issues like the communication between internal elements. In fact, classical buses could not assure a reliable connection between them. A new solution has to be found to face this problem. In 2002, the NoC paradigm has been introduced by Luca Bennini and Giovanni De Micheli [4]. This proposal has resolved the intra-communication problem and data exchange.

The NoC paradigm was important because it allowed design engineers to follow technology advancements and so, integrating many cores at the same chip by overcoming the intra-communication problems. For this purpose, many studies were conducted and many NoC architectures were proposed and later some of them were enhanced. We can find in the literature some relevant surveys and comparative studies between NoC proposals. Reference [5] details the NoC concept and discusses some examples. Other references also discusses this subject in many aspects by proposing a detailed comparison between NoC architectures and performances or by exposing the future of NoC related researches [6-8]. In our case, we will present the NoC concept to give lecturers an overview about it, this will be the subject of the second section. As we said before, this case study is focused on establishing a study about NoC dedicated tools. We will present our findings in this subject respectively in section 3 and 4. Finally, we review related works in section 5 and we conclude the work and expose perspectives in section 6.

## II. THE NoC CONCEPT

In this section we will introduce the NoC concept and later we will present some of their principal characteristics. At the end of this section we will show the research axes and the problems facing the research community when developing NoCs followed by some common NoC architecture proposals. In purpose to show the importance of NoC in recent SoCs, we decided to begin this section by a comparison between classical buses and Network-On-Chips. Table 1 gives a qualitative comparison between conventional buses and NoCs. As we can see this table demonstrates the usefulness of NoC based systems especially in case of the SoC contains many processing elements.

TABLE I. QUALITATIVE COMPARISON BETWEEN BUSES AND NOCS

| Buses | | | NoCs |
|---|---|---|---|
| Each item will add a parasitic capacitance, hence the degradation of electrical performance increase with the number. | - | + | The elements are connected by point to point interconnections for all sizes of networks, local performance is not degraded. |
| The time management of the bus is difficult. | - | + | The data transfer can be accomplished by delayed transitions because the connections are point to point. |
| Delays caused by arbitration at the bus can cause blockages, especially if the number of masters is important. | - | + | Routing decisions are distributed. |
| The bandwidth is limited and shared by all elements of the bus. | - | + | The bandwidth increases with the size of the network. |
| The bus tests are long and problematic. | - | + | The dedicated BIST (Built In Self Test) are locals, complete and fast. |
| The bus latency is the speed of a connection control circuit if the bus is granted. | + | - | Internal decisions making can add delays. |
| The concept is simple and easy to understand. | + | - | Designers need upgrades in order to exploit new concepts. |

*A. NoC Architecture*

The NoCs consist typically of routers, network adapter (network interface) and connections [6, 9].

*1) Router:* directs the data according to the protocol selected. It contains the routing strategy.

*2) Network Adapters:* provide a bridge between the router and the element attached to them. Their main task is to separate calculation (IPs) of the communication (network). This consists of two operations which are protocol conversion and packages construction.

*3) Connections:* are the channels of transmission of data between the various circuit elements to the network.

*B. Topology*

The topology of a network is the way in which routers, network adapters and connections are organized. There are several topologies that we can call regular or irregular [10-11]. This classification is based on the distribution of routers in the network. Figure 1 shows some regular topologies we can find a) mesh b) mesh torus c) ring d) fat-tree.

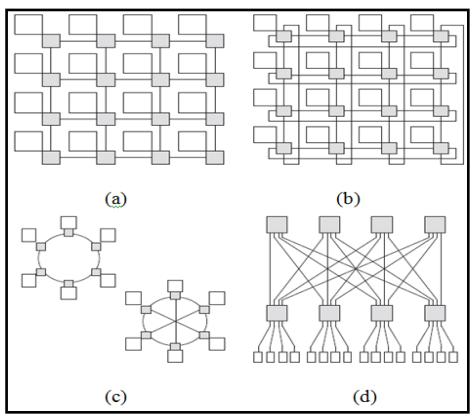

Fig. 1. Examples of regular NoC topologies

In the other side, irregular topologies are composed of two or three regular topologies such as a mesh topology and ring simultaneously.

*C. Routing*

Routing is to transfer data from source to destination with a clearly defined strategy. In the literature, researchers have classified the routing algorithms according to different criteria:

*1) The type routing is called source routing if only the sender provides the path by which the data will flow, it is called distributed if the transit decision is taken locally at each node. We can also find a classification similar to the previous one except that it defines a more general routing strategy regardless of source, in effect if the routing decisions are identically distributed across the network, routing is called centralized. If these decisions are taken locally, routing is still called distributed [5]. As we can see that decision does not take into account the sender as the one before.*

*2) The routing is deterministic if the transit path is determined by the sender and the receiver only. The path between the same network correspondents is invariable. However, if the transit of data between two network elements can be achieved through multiple paths, routing is then called adaptive. This is possible thanks to the decisions taken locally at the nodes. The implementation of adaptive routing algorithms can generate complicated nodes but can ensure a better flow of data within a NoC.*

*3) The routing is called circuit switching when a circuit (a path) between the transmitter and receiver is reserved for the duration required to transfer data. It is called packet routing when the data to be transmitted is divided into packets containing a portion of the data and routing information. The packets may follow different paths to reach their destination.*

A routing algorithm usually has one or more of the characteristics mentioned a little earlier. For example, an adaptive routing is generally a packet switching routing.

*D. Switching techniques*

The primary function of switches is to determine when and how the inputs of a router will be connected to its outputs [12]. There are several switching techniques among them store-and-forward, virtual cut-through and wormhole.

*1) Store-and-forward:* the transferred data is split into packets and each packet contains routing information. When a packet reaches a node, it is entirely saved in a buffer and routing information is extracted to determine the appropriate output port.

*2) Virtual cut-through:* the routing information is contained in the first bytes of the packet. Instead of saving the entire package like store-and-forward, the packages are sent as soon as the output port is determined. In case this port is used, the package will be saved in a buffer [6, 9].

*3) Wormhole:* the packets are split into sub-packets called flits (Flow Control Unit). The control data are contained in the header flit. As a result, a single packet can be transmitted by different nodes. This will reduce latency, but may cause many bottlenecks in the network.

*E. Related research axes and issues*

Scientific Research for NoCs is conducted on several axes. They can be classified into three broad categories or levels: networks, interconnection or system. In what follows we will describe these different orientations.

*1) Network Level*

The research at the network level is the most solicited level between scientific community. This is because NoC were in developing phases. The most discussed topics are the following:

*a) Topology:* Regular, non-regular or mixed.

*b) Protocol:* routing, switching

*c) Flow control of data:* Anti-blockage mechanism, virtual channels, buffering.

*d) Quality of Service (QoS):* throughput, latency

*2) Connection Level*

Interconnection level research can be considered as a direct consequence of that done at the network level. Since the interconnections are used to join the network adapters to routers and also routers between them, a non-optimized communication can affect network performances. So it is also important that interconnections have to be studied. The items concerned are:

*a)* Synchronization

*b)* Parallel vs. serial

*c)* Reliability

*d)* Pipeline

*3) System level*

Given the advances in research for NoC and especially at the architectural level (network), the current proposed NoCs are increasingly complicated. Not only conceptually but also at simulation and testing phases. Currently, we can find in the literature more than sixty proposals with very different configurations [6, 9]. This includes the topology, routing, switch mode, the implementation technology and even the method of simulation and evaluation. This mixture, added to an exponentially growing complexity of architecture, has prompted researchers to turn to new design methodologies. Methodologies where the level of abstraction is raised to the system level in order to facilitate the designers work. The research for NoCs at the system level is summarized in the following:

*a) Design methodology:* modeling, co-design.

*b) Evaluation and assessment of performance:* latency, throughput, power consumption and space.

*c) Architectures:* system-level composition, reconfigurable NoCs.

The advancements obtained from the researches at the system level have conducted to the development of many CAD tools dedicated for NoCs. These tools will allow the management of very complex NoCs architectures throughout the design flow (modeling, simulation and implementation). A second advantage came from the fact they are especially designed for NoCs and not for general use, thereby ensuring more relevant results. Before we develop this topic in details later in paragraph II and IV, we will present some NoCs proposals in what follows.

*F. Some NoCs proposals*

In this section we present examples of NoC architectures from the literature. We will restrict ourselves to some examples because the objective behind this research is not the architecture proposals but NoCs tools used on their development. They are many works that have focused on the collection, classification and characterization of different architectures and implementations of NoCs to date. For a more complete list of NoCs proposals, we advice lecturers to consult these references [5-9].

*1) ÆTHREAL*

This Network-On-Chip was developed by Philips Research Laboratories. This NoC offers QoS for data transfer within a SoC, such as a) No loss b) No corruption c) Organized Transfer Order and so the transfer rates are guaranteed and the latency is predictable [13-14].

*2) SPIN*

The SPIN architecture has been developed by the University "Pierre et Marie Curie" [10]. The main characteristics of this NoC are a) expanded tree topology b) Routing packet.

*3) QnoC*

This NoC is developed by the Israeli Institute of Technology [15]. It is based on a mesh topology that can be irregular and the wormhole as switching technique.

### III. NOC DEDICATED TOOLS

Recent SoCs typically contain a relatively complicated architecture with a large number of computing elements [1, 16]. This requires a NoC based design to ensure an optimum management of the transit of internal data [3, 9]. To facilitate the development of embedded systems containing a network on chip, several dedicated tools have been proposed. These initiatives are often presented by the scientific community through research teams. Nevertheless, there are some other proposals from the industry.

The dedicated NoC tools vary depending on the purpose for which they are developed. We can distinguish two main classes: synthesizers and simulators. Regarding synthesizers, points often discussed are the quality of generated architectures (space, energy consumption) and the level of abstraction for modeling NoCs, the higher the level of abstraction is, the higher the design and its correction are fast. Recent compilers are becoming more powerful and it exists some commercial versions like FlexNoc from Arteries [17], INOC [18] and The Tool Suite Works CHAIN from Silistix [19-20]. In the other side, two criteria are often addressed for simulators: the estimation of power dissipation and performance computing (throughput, latency, and reliability). These two criteria are crucial since NoC are an integral component of an embedded system. Because these systems are often subject to hard constraints of space, energy and execution time, a relevant estimation of the NoCs characteristics could be very helpful to the designer. We will show in the following NoCs synthesis or simulation tools we collected from the literature. We recall that this list is not exhaustive.

*A. NS-2*

NS-2 was first developed for prototyping and simulating ordinary computer networks. However, since NoCs shares many characteristics with classic networks, NS-2 was widely used by many NoC researchers to simulate NoCs [21-22]. Many NoC studies have used NS-2 as a simulation tool making it a reliable reference especially when comparing the performances of two different architectures [23-25]. Finally, NS-2 is an open source, discrete event driven simulator and developed in C++ and OTcl. These modularity and availability have facilitated its spreading between researchers.

*B. Noxim*

This tool has been proposed by the Computer Architecture team at the University of Catania [26]. It is developed in SystemC language. It allows the user to define a 2D mesh NoC architecture with various parameters including: 1) Network size 2) Buffers size 3) Packet size 4) Routing algorithm 5) Injection rate of packets. Noxim allows the evaluation of NoCs in terms of throughput, latency and power consumption.

*C. DARSIM*

DARSIM is a NoC simulator which was developed at the Massachusetts Institute of Technology (MIT). This tool allows the simulation of mesh NoC architectures of 2 and 3 dimensions. It offers a multitude of NoC simulation configurations with various parameters. This includes two generation modes of data generation:

*a) Trace-driven injection* which involves the injection of packets into the network and monitors their spatial and temporal evolution. Each injection contains the tracing parameters that are time constraints, the identifier of the stream, the packet size and possibly the injection frequency.

*b) MIPS Simulation mode:* each node can be configured as a MIPS (Microprocessor Without Interlocked Pipeline Stages) with its own memory. These are connected to the NoC MIPS and receiving/sending data from/to the network can be simulated cycle by cycle. Using these methods and a more detailed explanation of the simulator are presented in this reference [27].

*D. SunFloor - 3D SunFloor*

SunFloor is a support tool for NoC design. It can be used at earlier design phases to synthesize the most appropriate topology with these constraints as input (Model, Energy and Space, Design Objectives). From these data, SunFloor generates a system specification ready to be translated into comprehensive architecture, usually in SystemC language and by the intervention of a second tool which is xpipesCompiler [28-29].

SunFloor 3D is an extension of the later version. The main feature added is the generation of specifications for the future 3D Wafers [30]. Both versions were developed by the team of Prof. Giovanni De Micheli, a pioneer of research for NoCs with many publications in the NoC subject like a particular article [4] with over 1900 citations (Google Scholar statistics) and several books on this subject [31].

*E. ORION 2.0*

ORION 2.0 is the successor of the version proposed by a team from Princeton University in 2003 [32]. It is a simulator dedicated primarily to the estimation of power and space for NoCs architectures. Among the improvements compared to the first version we find the support for new semiconductor technology through models of transistors and capacitances upgraded from industry [33-34].

*F. NoC Emulation techniques*

There are other techniques for simulation and the On-Chip verification of NoCs like the emulation technique proposed by [35]. This technique allows the emulation of NoC architectures such ÆTHREAL or those generated by xpipesCompiler through a standard platform. This involves interfacing IPs capable of injecting or retrieving data to and from the emulated NoC [36].

*G. Other available NoC tools*

Several other tools dedicated to NoCs are also available. The following table summarizes the tools we collected in the literature. We stress at the fact that list is not exhaustive.

TABLE II. NOC TOOLS PROPOSALS

| # | Tool | Year | Team | References |
|---|------|------|------|------------|
| 1 | NS-2 | 1995 | DARPA and later Contributors | [22] |
| 2 | Noxim | 2010 | Catagne University | [26] |
| 3 | DARSIM | 2009 | MIT | [27] |
| 4 | SunFloor – 3D | 2006 - 09 | EPFL (swizerland) | [28-30] |
| 5 | ORION 1 et 2 | 2003 - 09 | Princeton University | [32-34] |
| 6 | INSEE | 2005 | Basque University (Spain) | [37] |
| 7 | ATLAS | 2005 | Federal University of Brazil | [38] |
| 8 | NOCIC | 2004 | Massachussetts University | [39] |
| 9 | Pestannna Environment | 2004 | Phillips Research Laboratories | [40] |
| 10 | PIRATE | 2004 | Polytechnique School of Milan | [41] |
| 11 | SUNMAP | 2004 | Stanford University | [42] |
| 12 | xpipesCompiler | 2004 | Bologne University – Stanford University | [43] |
| 13 | µSpider | 2004 | Bretagne Sud University | [44] |
| 14 | OCCN | 2004 | ST microelectronics | [45] |
| 15 | NoCGEN | 2004 | University of New South Wales | [46] |
| 16 | FlexNoC | - | ARTERIS | [17] |
| 17 | iNoC | - | - | [18] |
| 18 | The CHAIN works tool suite | - | Silistix | [19] |

## IV. NoC Evaluation Methods

NoC Evaluation is an important step to classify the proposed architecture among the others. There are 3 general criteria or metrics that are considered by the majority of the research community: i) area consumption ii) power consumption and iii) latency. There is other metrics which were reported in some other works like packet loss or wire length. The authors of reference [6] reported that in average 3 metrics are discussed in any NoC proposal, which is inadequate says the author. In our study the NoC tools we found are globally focused on these three metrics cited above but some of them may have extra auxiliary options and configurations. Table 3 lists the NoC tool we collected by their characteristics. It includes modeling, simulation, hardware synthesis and availability. Besides, the modeling process includes many options which are: the network size, the buffers size, the packets distribution, the routing algorithm, the packet injection ration, the selection strategy and finally the traffic distribution. The simulation process contains: the area consumption, the power consumption, the network throughput and the latency.

## V. Related Works

Many works were done in the NoC area since their appearance. The major recent proposals are essentially based on more sophisticated architectures offering diverse advantages among them the quality of service (QoS) [12] or globally asynchronous locally synchronous (GALS) architectures which resolve the clocking difference problems inside a SoC [47-48].

Technology advancements have also pushed researchers to reconsider their point of view about NoCs. Besides, some works have focused on developing 3D NoC architectures [49] and as we have seen in section 3, NoC tools developers have also anticipated these advancements by proposing tools that are dedicated for 3D NoC design and simulation [30]. However, other studies proposed different approach by adding the NoC concept to the bus one and so, keeping some data transfer to classical buses. The objective is often to reduce coasts in term of area and power consumption and of course without degrading the system performances in terms of throughput and latency [50].

Other researchers have applied an existing concept which is basically developed for SoCs to the Network-On-Chip one like the reconfigurability. The term of ReNoCs which means Reconfigurable NoCs is more and more developed inside the scientific community and as a result some initiatives were elaborated on this subject [51].

## VI. Final Remarks And Conclusion

In this survey we tried to focus on a subject concerning NoCs that somehow was not deeply studied in the literature. In this paper we presented the NoC concept and its importance in recent SoCs.

Then we presented the tools dedicated to their development which includes the modeling, simulation and implementation processes. We stress again at the fact that this list is not exhaustive but can represent an important number of nowadays available NoC tools. Meanwhile, when we are writing this manuscript some other tools emerged and that we don't hesitate to include like MCoreSim [52] or a flexible parallel simulator with error control [53].

TABLE III. NOC TOOLS CHARACTERISTICS

| # | Tool | Specification | | | | | | | | | | | Hardware Synthesis | Availability |
|---|---|---|---|---|---|---|---|---|---|---|---|---|---|---|
| | | *Modeling* | | | | | | | *Simulation* | | | | | |
| | | NS | BS | PD | RA | SS | PIR | TD | AC | PC | T | L | | |
| 1 | NS-2 | + | + | + | + | + | + | + | + | + | + | + | - | + |
| 2 | Noxim | + | + | + | + | + | + | + | - | + | + | + | - | + |
| 3 | DARSIM | + | + | + | + | + | + | + | - | - | + | + | - | - |
| 4 | SunFloor – 3D | + | - | - | - | - | - | - | - | + | + | + | + | - |
| 5 | ORION 2.0 | - | + | - | - | - | - | - | + | + | - | - | - | - |
| 6 | ATLAS | + | - | + | + | - | + | - | - | - | + | + | + | + |
| 7 | PIRATE | + | + | - | - | - | - | - | + | + | - | - | - | - |
| 8 | SUNMAP | + | + | - | - | - | - | - | - | + | + | + | + | - |
| 9 | µSpider | + | + | - | - | - | - | - | - | - | + | + | + | - |
| 10 | NoCGEN | - | + | - | - | - | - | - | - | - | + | + | + | - |
| 11 | FlexNoC | + | + | + | + | + | + | + | + | + | + | + | + | commercial |
| 12 | iNoC | + | + | + | + | + | + | + | + | + | + | + | + | commercial |
| 13 | The CHAIN works tool suite | + | + | + | + | + | + | + | + | + | + | + | + | commercial |

TABLE IV. NOMENCLATURE

| Modeling | Simulation |
|---|---|
| **NS: Network Size** | AC: Area Consumption |
| **BS: Buffers Size** | PC: Power Consumption |
| **PD: Packets Distribution** | T: throughput |
| **RA: Routing Algorithm** | L: Latency |
| **PIR: Packets Injection Ratio** | |
| **SS: Selection Strategy** | |
| **TD: Traffic Distribution** | |